# ON THE CONNECTIONS BETWEEN SURFICIAL PROCESSES AND STRATIGRAPHY IN RIVER DELTAS

MICHAEL J. PUMA[1,6], RACHATA MUNEEPEERAKUL[1], CHRIS PAOLA[2,3], ANDREA RINALDO[4,5], IGNACIO RODRIGUEZ-ITURBE[1]

[1]*Department of Civil & Environmental Engineering, Princeton University, Princeton, NJ, 08544, USA*
[2]*Department of Geology & Geophysics, St. Anthony Falls Laboratory, University of Minnesota, Minneapolis, MN, 55414, USA*
[3]*National Center for Earth-surface Dynamics, University of Minnesota, Minneapolis, MN, 55414, USA*
[4]*Laboratory of Ecohydrology, Faculté ENAC, École Polytechnique Fédérale, CH-1015 Lausanne, Switzerland*
[5]*Department IMAGE & International Centre for Hydrology "Dino Tonini", Universitá di Padova, via Loredan 20, I-35131, Padova, Italy*
[6]*Center for Climate Systems Research, Columbia University, NASA Goddard Institute for Space Studies, 2880 Broadway, New York, NY 10025, USA*

e-mail: mpuma@giss.nasa.gov

**Abstract:**

We explore connections between surficial deltaic processes (e.g. avulsion, deposition) and the stratigraphic record using a simple numerical model of delta-plain evolution, with the aim of constraining these connections and thus improving prediction of subsurface features. The model represents channel dynamics using a simple but flexible cellular approach, and is unique in that it explicitly includes backwater effects that are known to be important in low-gradient channel networks. The patterns of channel deposits in the stratigraphic record vary spatially due to variation in avulsion statistics with radial distance from the delta's source of water and sediment. We introduce *channel residence time* as an important statistical measure of the surface channel kinematics. The model suggests that the mean channel residence time anywhere within the delta is nicely described by a power law distribution showing a cutoff that depends on radial distance. Thicknesses of channel deposits are not uniquely determined by the residence time of channelization. The channel residence time distributions at given radial distances from the source are found to be approximately lognormally distributed, a finding consistent with the scale-dependent radial structure of channel deposits in the stratigraphic record.

Keywords: River delta, avulsion, stratigraphy, scale invariance, power law

## INTRODUCTION

Deltas are fragile geomorphic systems controlled by complex interactions among sediment supply, geomorphological fluvial processes, relative sea-level change, and marine energy conditions (Giosan and Bhattacharya 2005; Orton and Reading 1993; Overeem et al. 2005; Rinaldo et al. 1999). The sensitivity of deltaic systems to these processes and conditions is particularly important given the expected effects of climate change as predicted by the Intergovernmental Panel on Climate Change (Parry et al. 2007). Unfortunately, the intrinsic complexity of deltas has hindered progress on quantification of the connections among surficial processes and depositional architecture and, consequently, on our ability to model and manage them (Giosan and Bhattacharya 2005).

For a delta model, it is convenient to identify processes in three sections: the delta plain, the river mouth/shoreline, and the subaqueous delta slope (e.g. Orton and Reading 1993). The focus of this paper is on processes and deposition in the delta plain, which is often a region that is of great importance to society (highly populated and/or valuable agricultural regions). The morphology of the delta plain is controlled by fluvial dynamics, flood plain sedimentation, base level, and tectonics. Avulsion, which is the diversion of flow out of an established channel to a new channel on the existing floodplain, is a particularly important process (Slingerland and Smith 2004). Early modeling efforts that focused on avulsion include the work of Leeder, Allen, Bridge, and coworkers (Allen 1978; Bridge and Leeder 1979; Leeder 1978; Mackey and Bridge 1992; Mackey and Bridge 1995). These models simulate channel avulsions in alluvial systems to predict channel and floodplain deposition. The general strategy of these models is to estimate channel and floodplain deposition using empirical relationships.

Other quantitative analyses of deltas have focused on physically-based modeling (Syvitski and Daughney 1992; Syvitski and Hutton 2001), which is one of the most promising approaches to quantify the connection among the surficial processes and depositional architecture of deltas. Overeem et al. (2005) reviewed current models that use physically-based approaches. The processes of sediment transport and avulsion on the delta plain are represented in a variety of ways. For example, the 3D-Sedflux model (Overeem et al. 2005) uses the linear diffusion model originally described by Paola et al. (1992). However, Jerolmack and Paola (2007) argue that the linear diffusion model does not take into account the highly non-linear nature of large-scale sediment transport, which has been observed in field and experimental studies. Avulsion, which can be viewed as a threshold process resulting from the interaction of channels and their floodplains, is primarily responsible for this non-linear behavior and affects the large-scale dynamics of the system (Jerolmack and Paola 2007).

Syvitski et al. (2005) recently analyzed the distributary characteristics for 55 of the world's deltas and suggest that a power-law relationship holds between the number of distributary channels and a delta's gradient or the length of a river's main stem. More research needs to be done to verify these relationships, but, even if they do hold, no process-based model exists to predict accurately the number of distributary channels in a delta. In light of this challenge, the model in this paper simulates an end-member case, a delta with a single active channel that avulses across the delta plain (14 of the 55 deltas in Syvitski et al. (2005) have a single active channel). We develop the model to represent an input-dominated river-delta plain subject to constant water and sediment supply. The model builds upon the work of Jerolmack and Paola (2007), who developed a physically-based avulsion model that represents a minimal set of processes necessary to capture the overall characteristics of the system. The current model differs in treatment of details related to channelized flow, avulsion, and channel width. Our focus here is on developing measures of channel kinematics that we can then relate to quantitative statistical properties of the resulting stratigraphy. The model we develop is not intended to be a fully detailed representation of delta dynamics but rather a vehicle for exploring the relationship of statistical channel dynamics to statistics of channel stratigraphy.

## MODEL OF DELTA-PLAIN DYNAMICS

The domain is a two-dimensional Cartesian grid with length $L$ and width $W$, where the active channel originates at a fixed grid cell on the left boundary of the domain and terminates at the right end of the domain. Water and sediment transport are modeled numerically in one dimension for the active channel, where water and sediment are supplied at a constant rate to the fixed source cell. A no-flow boundary condition to water and sediment, which physically correspond to a topographic high, is imposed at the top and bottom of the domain. The size of the grid cells in the model is assumed to be larger than the channel belt. Channel-scale properties (e.g. braided/meandering flow, bars) are not explicitly modeled and results must be interpreted at the channel-bed scale (e.g., Jerolmack and Paola 2007; Sun et al. 2002). Given this grid size, cells that are not occupied by the active channel represent the distal floodplain.

The model discriminates between two types of sediment, bedload and suspended sediment. The suspended sediment is assumed to be responsible for floodplain sedimentation. Jerolmack and Paola (2007) analyzed in detail several different formations for floodplain sedimentation. They pointed out that floodplain deposition (and erosion) is not well understood over large space and time scales. We assume that floodplain sedimentation is spatially uniform, their simplest formulation, given the lack of quantitative knowledge related to floodplain dynamics. The dynamics of bedload sediment are restricted to the active channel and are modeled using a physically-based approach.

*Channelized Bankfull Flow and Bedload Sediment Transport*

Recent cellular approaches for modeling flow in river deltas have assumed that flow in channels during floods (bankfull flow) can be approximated as steady and uniform (Jerolmack and Paola 2007; Sun et al. 2002). However, since a river delta, by definition, involves flow into standing water, backwater effects on flow in the channel are potentially important for modeling river deltas (e.g., Wright and Parker 2005). A schematic representation of the flow condition is given in Figure 1A. We assume that flow is one-dimensional and quasi-steady in a rectangular channel as (e.g., Parker 2004; Wright and Parker 2005)

$$\frac{d}{dl_{ch}}\left[\frac{1}{2}\frac{Q_w^2}{B^2 H^2} + gH + g\eta\right] + gS_f = 0 \qquad (1)$$

where $H$ is the depth of flow in the channel [L], $B$ is channel width [L], $Q_w = BUH$ is the volumetric water discharge [L$^3$/T], $U$ is the mean velocity of flow [L/T], $g$ is gravity [L/T$^2$], $\eta$ is the elevation of the bed surface [L], $S_f$ is the friction slope [-], and $l_{ch}$ is the streamwise coordinate. Through conservation of water mass and streamwise momentum, the standard backwater formulation for steady and gradually varied flow is

$$\frac{dH}{dl_{ch}} = \frac{S - S_f}{1 - F_r^2} \qquad (2)$$

where $S$ is the slope of the bed [-] and $F_r$ is the Froude number of flow [-]. The friction slope is defined as $S_f = C_f F_r^2$, where $C_f$ is the dimensionless bed resistance coefficient. $C_f$ is obtained by assuming that the Manning-Stickler formulation for bed resistance holds (e.g., Parker et al. 1998; Sun et al. 2002). This assumption gives

$$C_f = \alpha_r^{-2}\left(\frac{H}{k_s}\right)^{1/3} \qquad (3)$$

where $\alpha_r$ is a dimensionless resistance coefficient and $k_s$ is a roughness height [L]. The bed is assumed to be flat (without bedforms), such that $k_s \approx n_k D$, where $n_k$ is a dimensionless parameter taken here to be equal to 2. Deltas, however, typically have dune-covered beds. Therefore, we also simulated the delta with a constant value of $C_f$ and found that model predictions of surficial dynamics and stratigraphy were not very sensitive to the bedform assumption.

A relation is also needed for the channel width, $B$, across which flow and bedload-sediment transport occur. Although one approach is to set the channel width indirectly by assuming a constant boundary shear stress (see extended discussion in Paola (2000)), this approach has only been applied to steady and uniform flow conditions (Jerolmack and Paola 2007; Parker et al. 1998; Sun et al. 2002). Since we have gradually varied flow, we use an empirical relation to compute $B$ for bankfull flow in a sand-bed river. The expression is given as (Parker 2004)

$$B = 0.274 D \left( \frac{Q_w}{\sqrt{RgD}D^2} \right)^{0.585} \quad (4)$$

where $D$ is the diameter of the sediment and $R$ is the submerged specific gravity of the sediment (given by $R = \rho_s/\rho - 1$, where $\rho_s$ and $\rho$ are the densities of the sediment and water, respectively). The constants in the above expression were computed based on data from Church and Rood (1983) (see Parker (2004) for details).

Conservation of bed-sediment mass is given by the standard Exner equation and is expressed as (e.g., Wright and Parker 2005)

$$(1-\lambda)\left( \frac{d\eta}{dt} + \frac{\sigma}{I_f} \right) = -\frac{dq_s}{dl_{ch}} \quad (5)$$

where $\lambda$ is the porosity of deposited sediment [-], $\sigma$ is the subsidence rate [L/T], $I_f$ is an intermittency factor [-], and $q_s$ is the volumetric sediment discharge per unit width [L/T]. The intermittency factor is a simple way to account for flow variability, for which it is assumed that the long-term sediment transport in the delta is controlled by 'characteristic' events (obtained from the frequency-magnitude distribution of floods) (Swenson et al. 2005; Wright and Parker 2005). During periods of low flow (time fraction $1 - I_f$), subsidence still occurs, but it is assumed that the river delta is not transporting significant amounts of sediment. For a more detailed discussion of $I_f$, the reader is referred to Paola et al. (1992), Swenson et al. (2005), and Wright and Parker (2005).

A constitutive relationship is needed to relate the sediment flux to parameters characterizing the sediment and channel bed, so we assume that a general form of the Meyer-Peter and Muller (1948) formulation holds. It is expressed as (Parker et al. 1998)

$$\frac{q_s}{\sqrt{RgD}} = \alpha_s \left(\tau^* - \tau_c^*\right)^n \tag{6}$$

where $\tau^*$ is the dimensionless Shields stress, $\tau_c^*$ is the critical Shields stress associated with the threshold of sediment motion, and $\alpha_r$ is a dimensionless sediment-transport coefficient. Finally, the dimensionless Shields stress is given by

$$\tau^* = \frac{C_f q_w^2}{RgDH^2} \tag{7}$$

*Channel Avulsion and Incision*

The methodology for channel incision and avulsion is related to the work of Jerolmack and Paola (2007) and Sun et al. (2002). An initial channel is incised one channel depth into the floodplain. The channel depth is dependent on location due to backwater effects, whereas a constant depth is used in the model of Jerolmack and Paola (2007). An avulsion occurs when the channel has aggraded to an amount proportional to bankfull depth above the surrounding topography (Jerolmack and Paola 2007; Sun et al. 2002; Tornqvist and Bridge 2002), which is an approach based on the data of Mohrig et al. (2000). In particular, a channel avulses from the current path to a different path, if the slope of the path from the current cell $(i,j)$ to a non-active neighbor cell $(i+k,j+p)$ is greater than the slope from the current cell (*i,j*) to the downstream active cell (*i+m,j+n*) by an amount proportional to bankfull depth, $\beta H^{(i,j)}$ (here $\beta$ equals 1). The neighbor cells include the three downstream and two cross-stream cells. This condition is expressed as (Sun et al. 2002)

$$\frac{\eta^{(i,j)} - \beta H^{(i,j)} - \eta^{(i+k,j+p)}}{l_{ch}^{(i,j)\to(i+k,j+p)}} > S^{(i,j)\to(i+m,j+n)} \qquad (8)$$

where *S* refers to slope. A simplified schematic of the avulsion condition for the cross-stream cells is shown in Figure 1B.

The model keeps track of two elevations, a (lower) bed elevation, $\eta$, and a (higher) levee elevation. The levee elevation, $\eta_{levee}$, is used only to compute incision depth. When an avulsion occurs, a new channel is created from the point of avulsion following the path of steepest decent. If the difference between $\eta_{levee}$ and $\eta$ is less than the flow depth from the previously active channel ($\beta H^{(i,j)}$), then the $\eta$ is reduced so that it is lower than $\eta_{levee}$ by $\beta H^{(i,j)}$, which in this case corresponds to one channel depth ($\beta = 1$). For all times when incision does not occur, changes in $\eta_{levee}$ follow those of $\eta$. We note that Jerolmack and Paola (2007) used the levee elevation of the active cell minus the bed elevation of the neighbor cells to determine if an avulsion occurs. Because this approach led to excessive switching between a limited number of flow paths in preliminary simulations, the avulsion criterion (Equation (8)) of Sun et al. (2002) is implemented in this paper.

### *Solution and Simulation Parameters*

We solve the backwater equation (Equation (2)) using a explicit predictor-corrector scheme, while the Exner equation is solved using a forward Euler approximation in time and an upstream-weighting scheme in space (e.g., Parker 2004).    We chose parameter values to represent a sand-bed river delta (Tables 1 and 2).  The initial slope, $S_0$, is set to $2\times10^{-4}$, and the downstream flow depth, $H_{ds}$, is set to 5 m.

## RESULTS AND DISCUSSION

The analyses are performed for a delta that has reached a statistically steady state of the dynamics, where the sediment supply to the delta approximately matches the subsidence (a condition that is met when the so-called capture ratio is approximately equal to one).  As previously described, the simulated delta has a single channel that avulses across the deltaic plain.  Figure 2A presents locations of the channel for three instances in time.  The fraction of the delta visited by the channel, $f_v$, as a function of time interval, $\Delta t$, is shown in Figure 2B.  Figure 2B also reveals that the time that it takes the channel to visit every cell in the domain, $T_{fv=1}$, is equal to 1150 yr.  We normalize time-related variables by this value, because it is a time scale that characterizes the surficial dynamics (to the extent that it is a measure of the time required for the channel to traverse the floodplain).  The topography of the channel bed at the simulation end time, $12T_{fv=1}$, is presented in Figure 2C, which shows that elevation decreases with increasing radial distance from the source.

Although the model simulates a single active channel, the sets of visited cells for some time intervals have the appearance of a distributary network.  Figure 3 shows the visited cells for increasing normalized time intervals, $\Delta t/T_{fv=1}$, with the normalized channel visit (or residence time), $T/T_{fv=1}$, indicated by the color bar.  The number of channels and the channel-residence times have a clear dependence on distance from the source. In Figures 3A and 3B, the number of channels increases from a single channel to multiple channels (5 and 6, respectively) at the downstream boundary of the model. The width of the distributary network also appears to increase farther from the source, with larger distances between adjacent channels.

Channels extend out from the source cell in all directions within the domain, which means that the flow direction near the source is often in the transverse rather than downstream direction. This behavior is demonstrated in Figure 3D, where a channel flows in the transverse direction down close to the bottom boundary ($y/L \approx 0.6$) and then continues downstream (because of the no-flow boundary condition at the bottom). However, the tendency for transverse flow decreases with increasing radial distance. The no-flow boundary conditions at the top and bottom of the domain also lead to relatively long channel residence times near the boundaries.

For longer time intervals in Figures 3C and 3D, several features of the surficial dynamics are evident. First, avulsion locations are not restricted to a relatively fixed area of the floodplain (e.g. nodal avulsions) but rather may occur anywhere along an active channel (referred to in the literature as random avulsion (Slingerland and Smith 2004)). It is also clear from Figures 3 that avulsions can be small-scale events that either rejoin or form a new channel close (within a few grid cells) to the original channel. Alternatively, an avulsion can be a large-scale event (a regional avulsion) that occupies a section of the delta far from the original channel (Heller and Paola 1996; Slingerland and Smith 2004).

We also find in Figures 3C and 3D that the range of channel residence times spans more than four decades. Only a few channels, which are the result of regional avulsions, extend the entire length of the delta for long residence times ($\Delta t / T_{fv=1} \sim \mathcal{O}(10^{-2})$).

These channels have a relatively high proportion of visited cells around them, because avulsions are more frequently small scale (within a few cells from the previous channel). These figures also demonstrate that channel residence time tends to decrease with radial distance from the source (excluding the cells near the top and bottom boundaries). However, we note an interesting feature in the upper-right portion of the domain. For two locations, we find a channel with a long residence time (blue/purple cells) that is not matched by any single upstream channel (green cells) with a similar residence time. This pattern demonstrates that new channels occasionally reoccupy previously abandoned channels.

### *Stratigraphy of Transverse Sections*

Figure 4 shows the resulting stratigraphy in the transverse direction (including deposits before steady-state conditions were reached) at four locations downstream from the source. We can distinguish between two types of channels deposits for the sections close to the source, $0.04L$ and $0.2L$ (where $L$ is the downstream length of the domain), in Figures 4A and 4B. The transverse-flow deposits are the continuous layers of channel deposits with the elevation of the connected deposits decreasing with distance from the section center. Deposits from downstream flow are generally in the middle of the sections ($y/L$ ranging from approximately -0.15 to 0.15 and -0.2 to 0.2, respectively).

To focus on the downstream-flow deposits, we present the stratigraphy at $0.5L$ and $0.8L$ in Figures 4C and 4D, respectively, where there are no transverse-flow deposits. These figures demonstrate a tendency for clustering of channel deposits. That is, a given channel deposit often has an adjacent channel deposit, which has been described as 'connectedness' by Leeder (1978) and Bridge and Mackey (1993). An important distinction when assessing the channel deposits is whether the connectedness is the result of successive or non-successive occupation of adjacent cells. For deposits that are adjacent but were not formed due to successive occupation, the treatment of floodplain sedimentation becomes critical for determination of channel-deposit connectedness (which should be the focus of future investigations). If we compare the topography of the sections at $0.5L$ and $0.8L$ in Figures 5A and 5B, we find that the difference in elevation between local minima and maxima is generally greater for the section farther from the source (0.8L). This increase in topographic roughness downstream reduces the opportunity for successive occupation of cells (Fig. 5) and, consequently, has the potential to affect the connectedness of deposits.

The topographic roughness described above is related to the probability of an avulsion occurring upstream of a given transverse section. This probability increases with distance from the source, which implies that a downstream location has the potential to be more frequently affected by avulsions. That is, channels farther downstream should be abandoned more frequently than upstream locations, unless the majority of avulsions lead to a channel rejoining its parent channel downstream. The visited cells and channel residence times in Figure 3 confirm this behavior. Channel reoccupation does occur in the model, as discussed for Figure 3D, but it only tends to happen far downstream of an avulsion, where the bed elevation of an abandoned channel could be lower than the adjacent floodplain. That is, abandoned channels tend to act as repellers for reoccupation close to the avulsion location, attractors for reoccupation farther downstream (if they are local minima), and neutral sites otherwise.

The normalized histograms of the channel-deposit thicknesses for three transverse sections are shown in Figure 6. The thickness histograms are obtained by considering the thickness of each channel layer for all of the cells in the section. Since the flow depth, $H$, is greater in downstream locations due to backwater effects, a channel must be elevated above the adjacent floodplain by a greater amount than for upstream locations for an avulsion to occur. However, the histograms in Figure 6 shift to the left with increasing distance from the source, which indicates thinner channel layers. Therefore, we can infer from this shift that downstream channels are often abandoned—due to upstream avulsions—before they have aggraded to one channel depth.

### *Channel Residence Time*

We propose that a basic measure for quantitatively relating surficial processes to their associated stratigraphic record is the channel residence time, $T$. In terms of relating $T$ to the stratigraphy, we might expect that the depth of channel deposits, $h_{ch}$, should be proportional to the channel residence time. In Figure 7A, a scatter plot shows the total depth of channel deposits, $\Sigma h_{ch}$, versus the normalized total channel residence time, $\Sigma T / T_{f_v=1}$, for each cell in the domain over the entire simulation. We find that no deterministic relationship exists between depth of channel deposits and channel residence time. The relationship is highly non-unique, meaning that a distribution of $\Sigma h_{ch}$ values exist for a given value of $\Sigma T / T_{f_v=1}$. We consider the normalized histograms of $\Sigma h_{ch}$ for three ranges of $\Sigma T / T_{f_v=1}$ in Figure 7B, because there is significant scatter in the relationship. The histograms of $\Sigma h_{ch}$ shift to the right, which demonstrates that longer residence times tend to have thicker deposits. The histogram shapes also change from an approximately symmetric distribution for the shortest time to skewed distributions for the longer residence times.

We next consider the mean channel residence time, $\bar{T}$, for each cell by recording the duration of each visit and then averaging all of these channel visits. A subsection of the *LxW* domain (with size *LxL*) is used for this analysis, so that the effects due to the no flow boundaries at the top and bottom of the domain are minimized. The mean channel residence time is a quantitative characterization of mass transport in a delta and the space-time coverage of the deltaic plain by a channel. Figure 8A shows a clear dependence on radial distance from the source for the normalized mean channel residence time, $\bar{T}/T_{f_v=1}$, with closer location having larger values. We are interested in properties that exhibit scale invariance in deltaic processes. In this case, we investigate to see if the distribution function of $\bar{T}/T_{f_v=1}$ obeys a power law,

$$P\left[\bar{T}/T_{f_v=1} \geq t\right] \propto t^{-\chi} \tag{9}$$

where $t$ is an arbitrary value of the random variable $\bar{T}/T_{f_v=1}$. The probability distribution $P\left[\bar{T}/T_{f_v=1} \geq t\right]$ is plotted in Figure 8B, which shows that the distribution of $\bar{T}/T_{f_v=1}$ shows excellent agreement with a power law form over approximately two decades. Significantly, the power-law exponent $\chi$, equal to 1.7±0.3, does not change when the size of the domain is increased (along with sediment flux). The largest simulated domain is 10,000 m × 14,000 m, with a maximum *LxL* subsection of 10,000 m × 10,000 m.

This finding of scale invariance in the mean channel residence time bears important consequences on the relation between the surficial properties of a river delta and its stratigraphic record. In fact, Jerolmack and Paola (2007) suggested that avulsions in a river delta system are analogous to earthquakes, in that crossing an avulsion threshold releases stored sediment in the same way that an earthquake releases stored strain energy. These authors found some evidence of a power law by considering the distribution of waiting time between avulsions (anywhere in the domain), but their distributions had exponential tails, which they attributed to finite size-effects.

We further investigate the delta's scaling properties by analyzing the distributions of normalized residence times, $T/T_{f_v=1}$, rather than the mean values. For this case, we need to consider that $T/T_{f_v=1}$ is statistically non-stationary in space and hypothesize that $T/T_{f_v=1}$ is a function of radial distance from the source. The normalized histograms of $T/T_{f_v=1}$ are obtained by considering all of the $T/T_{f_v=1}$ values for cells whose center lies within the annulus between $R$ and $R+\Delta R$ within the subsection *LxL*. The distributions from the individual cells were first compared with the ensemble distribution (all $T/T_{f_v=1}$ values from all the cells within the annulus) and did not differ considerably from the ensemble histogram. The radial-distance-specific ensemble normalized histograms of $T/T_{f_v=1}$ are shown in Figure 9 and are approximately lognormally distributed. They shift to the left with increasing distance from the source, implying shorter residence times.

We found that locations with longer channel residence times tend to have thicker deposits for the entire delta (Fig. 7). The common dependence of channel residence time (Fig. 9), topographic roughness (Figs. 4 and 5), and thickness of channel deposits (Fig. 6) on distance from the source further demonstrates the connection between surficial dynamics and the structure of the stratigraphy. That is, the surficial dynamics of the delta, quantified as channel residence time, affect the thickness and connectedness of channel deposits. Consequently, future efforts to understand the connection between surficial dynamics and the stratigraphic record will potentially benefit from analysis of channel residence time.

## CONCLUSIONS

Several interesting findings have been revealed with this simple river delta model that relate stratigraphy with surficial processes. Both the channel residence time and stratigraphy are dependent on distance from the water and sediment source. The distributions of channel residence times at given radial distances from the source vary and are approximately lognormally distributed, while the mean channel residence time for any location in the delta is nicely described by a power law distribution. The spatial dependence of channel residence time is evident in the 'connectedness' and thickness of transverse-section stratigraphy. A major obstacle to the ultimate goal of relating surficial processes and stratigraphy is the finding that the duration of a channel visit to a location does not uniquely determine the thickness of the deposits. It would be useful for future modeling investigations to further quantify the distributions of channel deposit thickness as a function of channel residence time and to consider the spatial dependence of these distributions. In addition, various representations of floodplain sedimentation need to be explored in terms of their effects on these statistical properties of channel deposits.

## ACKNOWLEDGEMENTS

The authors gratefully acknowledge support by the STC program of the National Science Foundation via the National Center for Earth-surface Dynamics under agreement Number EAR- 0120914. We also thank Jan Nordbotten, Matthew Wolinsky, and Doug Jerolmack for their insightful discussions and comments.

## REFERENCES

ALLEN, J.R.L., 1978, Studies in fluviatile sedimentation: An exploratory quantitative model for the architecture of avulsion-controlled alluvial suites: Sedimentary Geology, v. 21, p. 129-147.

BRIDGE, J.S., and LEEDER, M.R., 1979, Simulation-Model of Alluvial Stratigraphy: Sedimentology, v. 26, p. 617-644.

BRIDGE, J.S., and MACKEY, S.D., 1993, A revised alluvial stratigraphy model, *in* Marzo, M., and Puigdefabregas, C., eds., Alluvial sedimentation, International Association of Sedimentologists Special Publication 17, p. 319-336.

CHURCH, M., and ROOD, K., 1983, Catalogue of Alluvial Channel Regime Data: Vancouver, BC, University of British Columbia, Department of Geography.

GIOSAN, L., and BHATTACHARYA, J.P., 2005, New directions in deltaic studies, *in* Giosan, L., and Bhattacharya, J.P., eds., River deltas : concepts, models, and examples: SEPM special publication: Tulsa, Oklahoma, Society for Sedimentary Geology, p. 502.

HELLER, P.L., and PAOLA, C., 1996, Downstream changes in alluvial architecture: an exploration of controls on channel-stacking patterns: Journal of Sedimentary Research, v. 66, p. 297-306.

JEROLMACK, D.J., and PAOLA, C., 2007, Complexity in a cellular model of river avulsion: Geomorphology, v. 91, p. 259-270.

JEROLMACK, D.J., and PAOLA, C., In press, Complexity in a cellular model of river avulsion: Geomorphology.

LEEDER, M.R., 1978, A quantitative stratigraphic model for alluvium, with special reference to channel deposit density and interconnectedness *in* Miall, A.D., ed., Fluvial Sedimentology, Canadian Society of Petroleum Geologists, Memoir 5, p. 587-596.

MACKEY, S.D., and BRIDGE, J.S., 1992, A Revised Fortran Program to Simulate Alluvial Stratigraphy: Computers & Geosciences, v. 18, p. 119-181.

MACKEY, S.D., and BRIDGE, J.S., 1995, 3-Dimensional Model of Alluvial Stratigraphy - Theory and Application: Journal of Sedimentary Research Section B-Stratigraphy and Global Studies, v. 65, p. 7-31.

MEYER-PETER, E., and MULLER, R., 1948, Formulas for bedload transport: Proceedings of the 2nd Congress of International Association for Hydraulic Research, p. 39-64.

MOHRIG, D.C., HELLER, P.L., PAOLA, C., and LYONS, W.L., 2000, Interpreting avulsion process from ancient alluvial sequences: Gaudeloupe-Matarranya system, northern Spain, and Wasatch Formation, western Colorado: Geological Society of America Bulletin v. 112, p. 1787-1803.

ORTON, G.J., and READING, H.G., 1993, Variability of Deltaic Processes in Terms of Sediment Supply, with Particular Emphasis on Grain-Size: Sedimentology, v. 40, p. 475-512.

OVEREEM, I., SYVITSKI, J.P.M., and HUTTON, E.W.H., 2005, Three-dimensional numerical modeling of deltas, in Giosan, L., and Bhattacharya, J.P., eds., River deltas : concepts, models, and examples: SEPM special publication: Tulsa, Oklahoma, Society for Sedimentary Geology, p. 502.

PAOLA, C., 2000, Quantitative models of sedimentary basin filling: Sedimentology, v. 47, p. 121-178.

PAOLA, C., HELLER, P.L., and ANGEVINE, C.L., 1992, The large-scale dynamics of grain-size variation in alluvial basins, 1: Theory: Basin Research, v. 4, p. 73-90.

PARKER, G., 2004, 1D Sediment Transport Morphodynamics with Applications to Rivers and Turbidity Currents (E-book).

PARKER, G., PAOLA, C., WHIPPLE, K.X., and MOHRIG, D., 1998, Alluvial fans formed by channelized fluvial and sheet flow. I: Theory: Journal of Hydraulic Engineering-Asce, v. 124, p. 985-995.

PARRY, M.L., CANZIANI, O.F., PALUTIKOF, J.P., VAN DEN LINDEN, P.J., and HANSON, C.E., 2007, Climate Change 2007: Climate Change Impacts, Adaptation and Vulnerability. Contribution of Working Group II to the Fourth Assessment Report of the Intergovernmental Panel on Climate Change: Cambridge, UK, Cambridge University Press, p. 1000.

RINALDO, A., FAGHERAZZI, S., LANZONI, S., MARANI, M., and DIETRICH, W.E., 1999, Tidal Networks 2. Watershed Delineation and Comparative Network Morphology: Water Resour. Res., v. 35, p. 3905-3917.

SLINGERLAND, R., and SMITH, N.D., 2004, River avulsions and their deposits: Annual Review of Earth and Planetary Sciences, v. 32, p. 257-285.

SUN, T., PAOLA, C., PARKER, G., and MEAKIN, P., 2002, Fluvial fan deltas: Linking channel processes with large-scale morphodynamics: Water Resources Research, v. 38.

SWENSON, J.B., PAOLA, C., PRATSON, L., VOLLER, V.R., and MURRAY, A.B., 2005, Fluvial and marine controls on combined subaerial and subaqueous delta progradation: Morphodynamic modeling of compound-clinoform development: Journal of Geophysical Research-Earth Surface, v. 110, p. 1-16.

SYVITSKI, J.P.M., and DAUGHNEY, S., 1992, Delta2 - Delta-Progradation and Basin Filling: Computers & Geosciences, v. 18, p. 839-897.

SYVITSKI, J.P.M., and HUTTON, E.W.H., 2001, 2D SEDFLUX 1.0C: an advanced process-response numerical model for the fill of marine sedimentary basins: Computers & Geosciences, v. 27, p. 731-753.

SYVITSKI, J.P.M., KETTNER, A.J., CORREGGIARI, A., and NELSON, B.W., 2005, Distributary channels and their impact on sediment dispersal: Marine Geology, v. 222, p. 75-94.

TORNQVIST, T.E., and BRIDGE, J.S., 2002, Spatial variation of overbank aggradation rate and its influence on avulsion frequency: Sedimentology, v. 49, p. 891-905.

WRIGHT, S., and PARKER, G., 2005, Modeling downstream fining in sand-bed rivers. I: formulation: Journal of Hydraulic Research, v. 43, p. 613-620.

**Table 1.** Parameters characterizing the bedload sediment and its transport properties (Parker et al. 1998; Sun et al. 2002).

| Parameter | Value |
|---|---|
| Characteristic grain size, $D$ | $3 \times 10^{-4}$ m |
| Submerged specific gravity, $R$ | 1.65 |
| Deposited-sediment porosity, $\lambda$ | 0.4 |
| Resistance coefficient, $\alpha_r$ | 15 |
| Roughness height, $k_s$ | $2D$ |
| Critical Shields stress, $\tau_c^*$ | 0 |
| Exponent in the sediment-transport relation, $n$ | 2.5 |
| Sediment-transport coefficient for a straight flume, $\alpha_{so}$ | 11.25 |
| Sediment-transport coefficient to account for meandering/braided channels, $\alpha_{sa}$ | 1.5 |

**Table 2.** Parameters used for the river delta simulation.

| Parameter | Value |
|---|---|
| Length of domain, $L$ | 5000 m |
| Width of domain, $W$ | 7000 m |
| Cell length & width, $\Delta L$ & $\Delta W$ | 100 m |
| Maximum time increment, $\Delta t_{max}$ | $2 \times 10^{-3}$ yr |
| Volumetric water feed rate, $Q_w$ | 20 m³/s |
| Volumetric sediment feed rate, $Q_s$ | $2.5 \times 10^{-3}$ m³/s |
| Flood intermittency factor, $I_f$ | 0.05 |
| Subsidence rate, $\sigma$ | $1 \times 10^{-3}$ m/yr |
| Floodplain aggradation rate, $\nu_{fp}$ | $0.8\sigma$ |
| Gravitational acceleration, $g$ | 9.81 m²/s |
| Depth of flow at downstream boundary, $H_{ds}$ | 5 m |

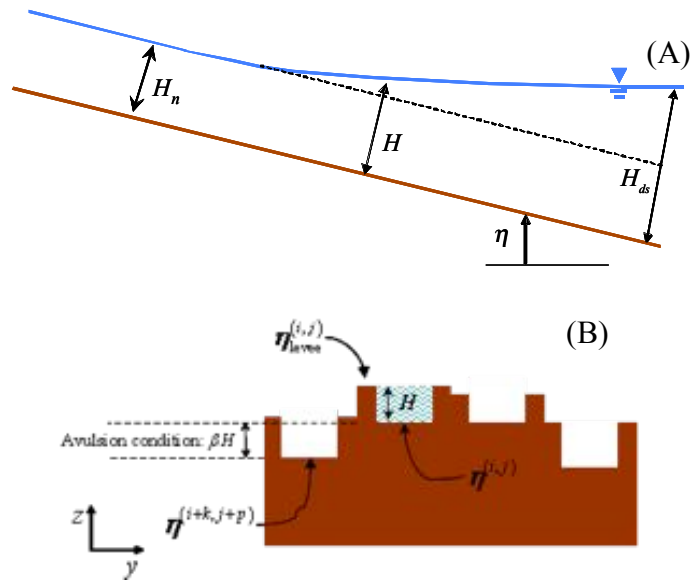

FIG. 1 – (A) Schematic representation of flow depth in the channel, *H*, as affected by the downstream water depth, $H_{ds}$. $H_n$ represents the flow depth for uniform, steady-state (normal) conditions. (B) Schematic of the condition for a cross-stream avulsion from (*i+m,j+n*) to (*i+k,j+p*). An avulsion will occur when the channel bed is elevated above a neighboring cell (3 downstream and 2 cross-stream cells) by a height that is proportional to bankfull depth, *βH*.

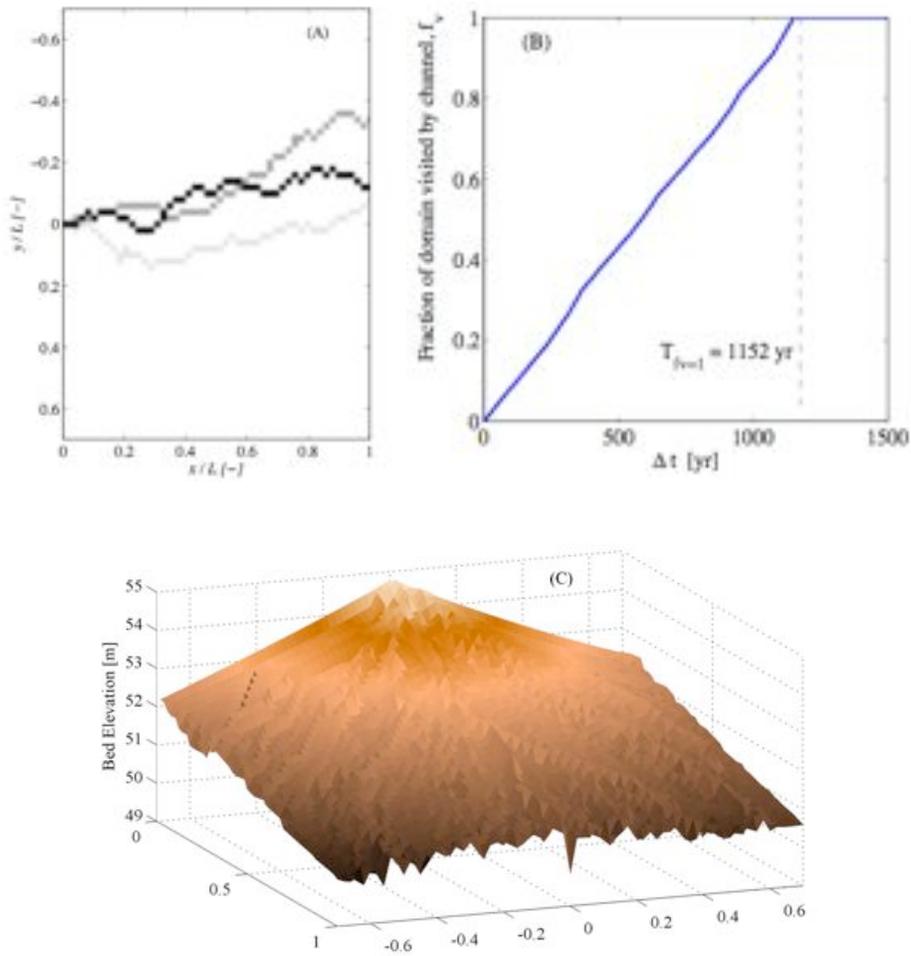

FIG. 2 – (A) Snapshots of active channel at three different times, (B) fraction of delta visited by the channel, $f_v$, as a function of time, and (C) final bed elevation, $\eta$, of the simulated delta plain, $L \times W$, at time $T_{max} \approx 12 T_{fv=1}$.

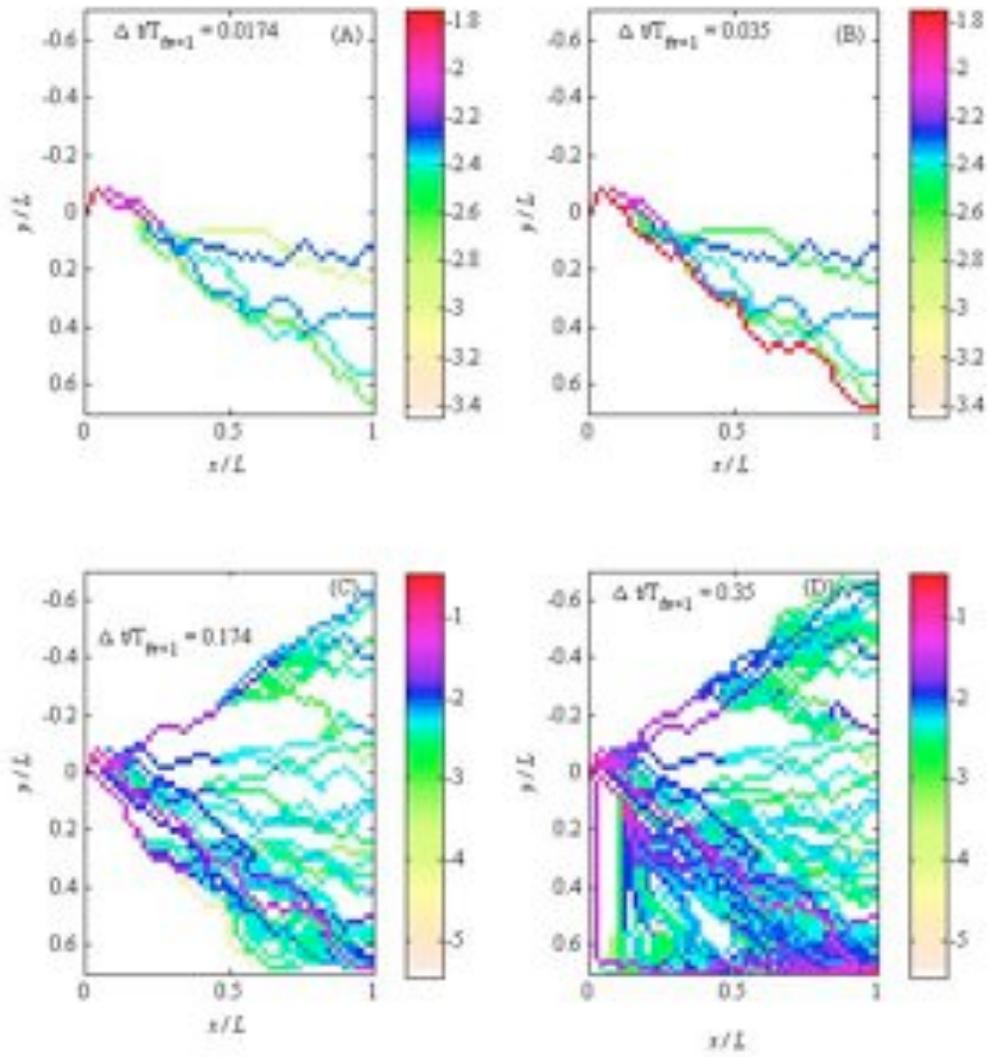

FIG. 3 – Visited cells for time intervals, $\Delta t/T_{f_v=1}$, equal to (A) 0.0174, (B) 0.035, (C) 0.174, and (D) 0.35 with the base-10 logarithm of normalized channel residence time, $T/T_{f_v=1}$, indicated by the color bar.

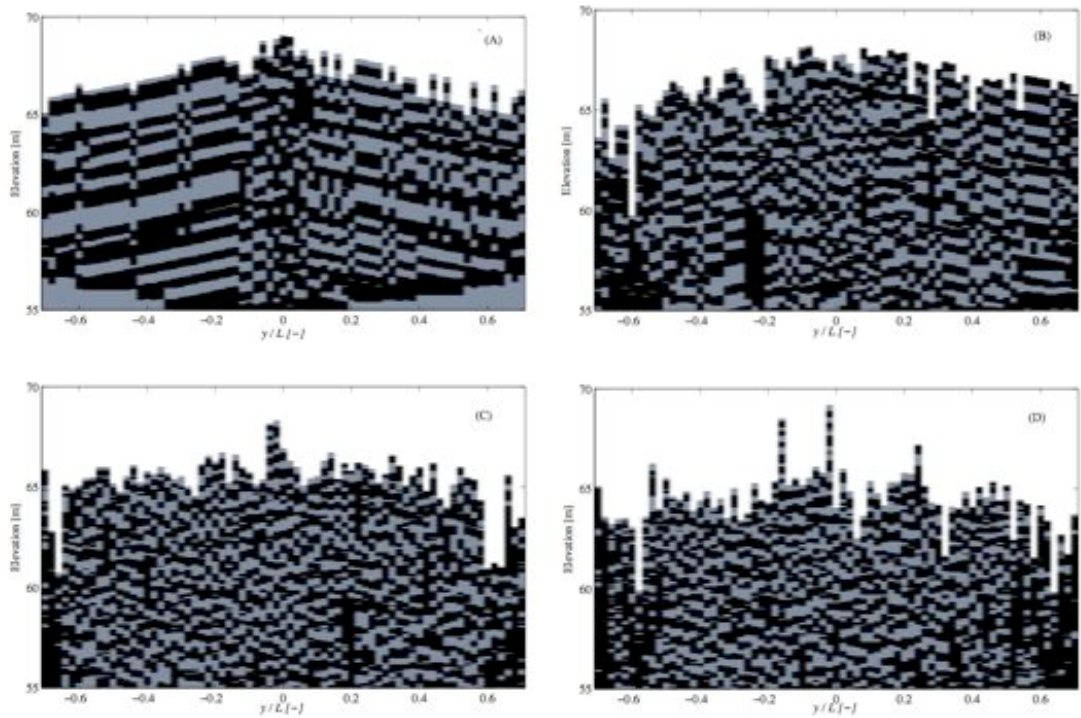

FIG. 4 – Stratigraphy of transverse sections with increasing downstream distance, *x*, from the water/sediment source: (A) 0.04*L*, (B) 0.2*L*, (C) 0.5*L*, and (D) 0.8*L*. The channel deposits are black and floodplain deposits are gray.

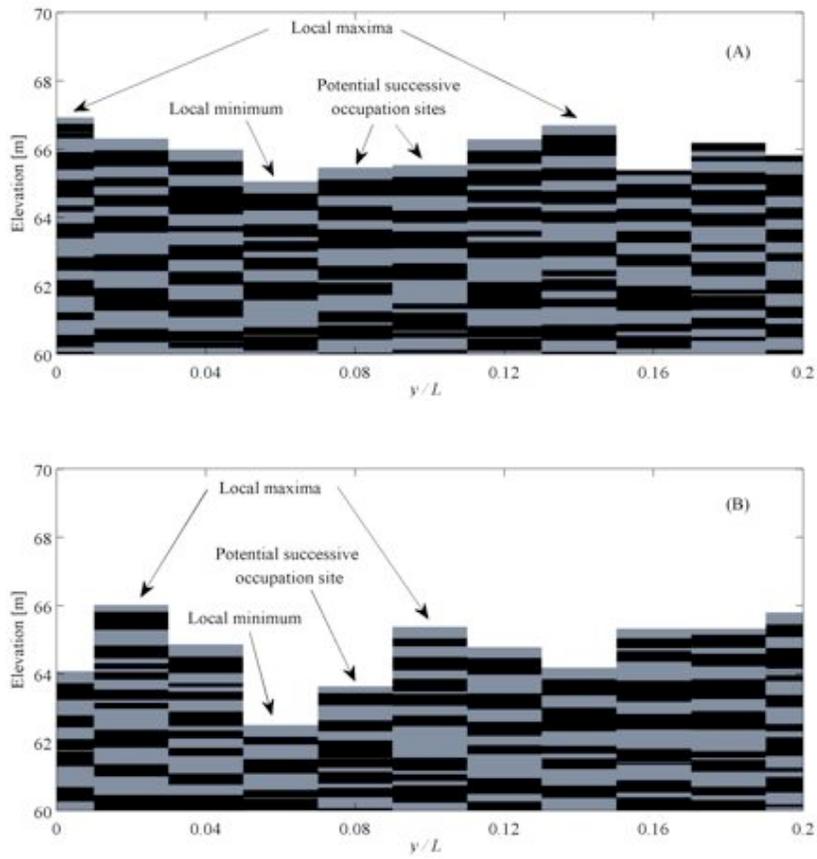

FIG. 5 – Stratigraphy of surface, identifying local topographic minima and maxima for the transverse sections at downstream distances (A) 0.5$L$ and (B) 0.8$L$.

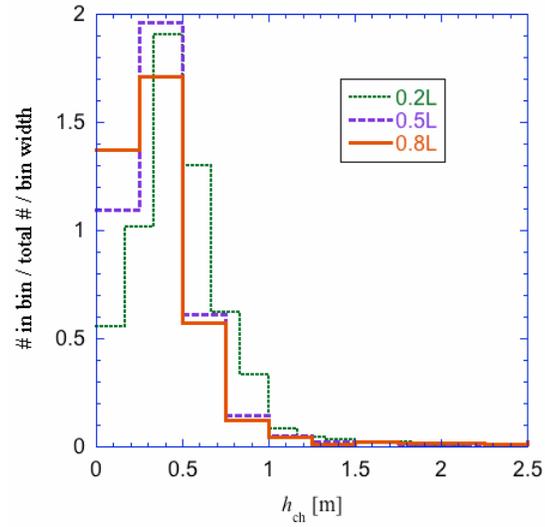

FIG. 6 – Normalized histograms of the thickness of each channel deposit layer for three transverse sections (0.2*L*, 0.5*L*, and 0.8*L*). Thicknesses were computed for each column in the transverse direction.

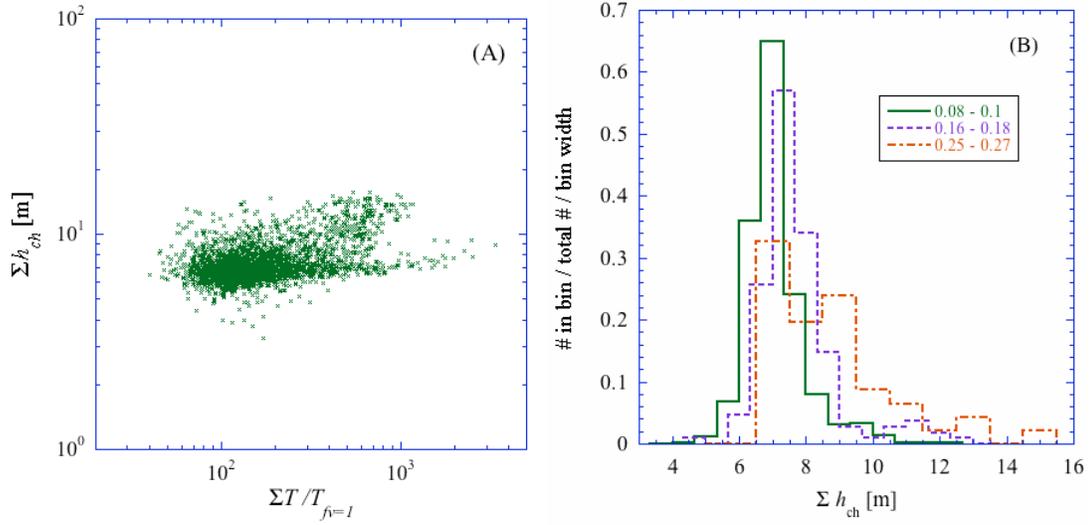

FIG. 7 – (A) Scatter plot of the normalized total time of channel residence, $\sum T / T_{f_v=1}$, versus the total depth of channel deposits, $\Sigma h_{ch}$, for each grid cell in the domain; (B) normalized histogram of $\Sigma h_{ch}$ for three ranges of $\sum T / T_{f_v=1}$.

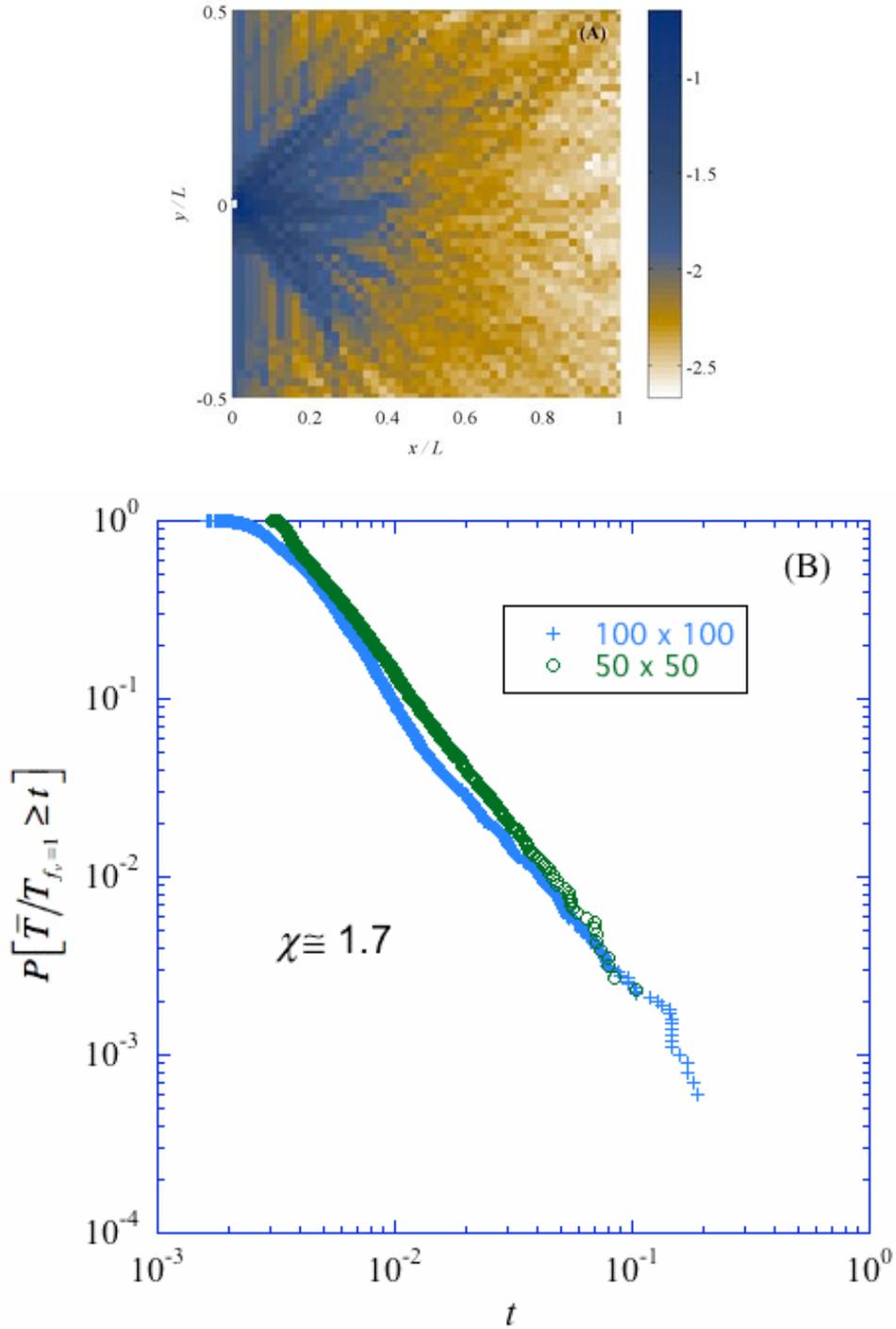

FIG. 8 – (A) Base-10 logarithm of normalized mean channel residence time, $\bar{T}/T_{f_v=1}$, for all cells within subsection $L$x$L$; (B) exceedance probability plots of the $\bar{T}/T_{f_v=1}$ values from (A) and for a larger domain.

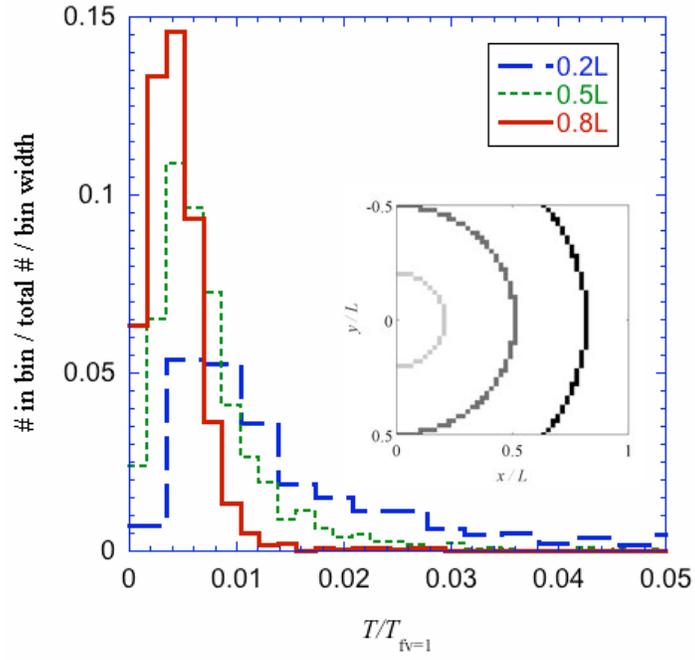

FIG. 9 – Normalized histograms of $T/T_{f_v=1}$ for each cell whose center lies within the annulus between $R$ and $R+\Delta R$ for three average radial distance, $\bar{R} = (R+\Delta R)/2$ : 0.2$L$, 0.5$L$, and 0.8$L$.